\documentclass[twocolumn,showpacs,preprintnumbers]{revtex4}
\usepackage{amssymb}
\usepackage{amsmath}
\usepackage{graphicx}
\usepackage{dcolumn}
\usepackage{bm}
\usepackage{color}

\begin{document}

\title{Quantum criticality of the Rabi-Stark model at finite frequency ratios%
}
\author{ Xiang-You Chen$^{1}$, You-Fei Xie$^{1}$, and Qing-Hu Chen$^{1,2,*}$}

\address{
$^{1}$ Zhejiang Province Key Laboratory of Quantum Technology and Device, Department of Physics, Zhejiang University, Hangzhou 310027, China \\
$^{2}$  Collaborative Innovation Center of Advanced Microstructures, Nanjing University, Nanjing 210093, China
 }
\date{\today }

\begin{abstract}
In this paper, we analyze the quantum criticality of the Rabi-Stark model at
finite ratios of the qubit and cavity frequencies in terms of the energy
gap, the order parameter, as well as the fidelity, if the Stark coupling
strength is the same as the cavity frequency. The critical exponents are
derived analytically. The energy gap and the length critical exponents are
different from those in the quantum Rabi model and the Dicke model. The
finite size scaling analysis for the order parameter and the fidelity
susceptibility is also performed. The universal scaling
behaviors are demonstrated and several finite size exponents can be then
extracted. Furthermore, universal critical behavior can be also established
in terms of the bosonic Hilbert space truncation number, and the
corresponding critical scaling exponents are found. Interestingly, the
critical correlation length exponents in terms of the photonic truncation
number as well as the equivalently effective length scales are different in
the Rabi-Stark model and the quantum Rabi model, suggesting they belong
to different universality classes. The second-order quantum phase transition
is convincingly corroborated in the Rabi-Stark model at finite frequency
ratios, by contrast, it only emerges at the infinite frequency ratio in the
original quantum Rabi model without the Stark coupling.
\end{abstract}

\maketitle
\section{Introduction}

As advocated by Anderson that more is different ~\cite{anderson}, emergent
phenomena occur when systems with many particles behave differently than
their original few ones. In classical systems the transition between
different phases is driven by thermal fluctuations. Similarly, quantum
fluctuations can lead to transitions between distinct quantum phases of
matter, such as superconductors and insulators ~\cite{Sachdev}. Recently,
the quantum Rabi model (QRM)~\cite{Rabi} and Jaynes-Cummings model~\cite{JC}%
, both only describe a single-mode cavity field and a two-level atom, but
interestingly exhibit the continuous quantum phase transition (QPT) \cite%
{plenio2015,plenio2016,MXLiu}, thus extending the previous concept to
systems with few degrees of freedom. To see the collective behaviors, one
need enhance the atom-field interaction and increase the two level energy
difference at the same time. The former requirement can be gradually
satisfied with the recent progress on cavity quantum electrodynamics (QED) ~%
\cite{CQED}, circuit QED ~\cite{Niemczyk,Yoshihara,Forn2}, and trapped ions~%
\cite{Leibfried}, which can realizes the QRM from strong coupling~\cite%
{Clarke}, ultra-strong coupling ~\cite{Forn}, and even to deep-strong
coupling ~\cite{Casanova}.

The light-matter interaction can be engineered in many different ways.
Benefiting from development of quantum simulation technology, the so-called
quantum Rabi-Stark model (RSM) has been realized by adding an nonlinear
process to the QRM \cite{Grimsmo1,Grimsmo2}, and the Hamiltonian is given by
\begin{equation}
H_{R}=\left( \frac{\Delta }{2}+Ua^{\dagger }a\right) \sigma _{z}+\omega
a^{\dagger }a+g(a^{\dagger }+a)\sigma _{x},  \label{QRSM}
\end{equation}%
where $\Delta $ and $\omega $ are frequencies of two-level system and
cavity, $\sigma _{i}(i=x,y,z)$ are usual Pauli matrices describing the
two-level system, $a$ ($a^{\dagger }$) are the annihilation (creation)
bosonic operators of the cavity mode with frequency $\omega $, and $g$
denotes the linear coupling strength between the qubit and the cavity. The
nonlinear coupling strength $U$ is determined by the dispersive energy
shift. To me more concise, the units are taken of $\omega =1$ throughout
this paper unless specified. It has been studied by the Bargmann approach~%
\cite{Maciejewski,Eckle} and the more physical Bogoliubov operator approach ~%
\cite{Xie2019}. It is found that the emergent Stark-like nonlinear
interaction bring about the novel and exotic physical properties. Such as,
the interaction induced energy spectra collapse can be observed in the limit
of $U\rightarrow \pm 1$ ~\cite{Xie2019} above a critical atom-cavity
coupling strength. The first-order QPT, indicated by the level crossing of
the ground-state and first excited state, is also observed~\cite{Xie2019}.
Grimsmo and Parkins conjectured that the nonlinear dispersive-type coupling
would possibly induce a new superradiant phase at this single atom level if
\ $U<-1$ ~\cite{Grimsmo1} in the open systems.

The RSM has been attracted much attention more recently. The implementation
of the RSM was proposed using trapped ions~\cite{Cong}. In this proposal,
external laser beams are applied to induce an interaction between an
electronic transition and the motional degree of freedom, thus the Stark
term is generated. A continuous quantum phase transition in the limit of $%
U\rightarrow \pm 1$ is suggested in terms of the energy gap closing at a
critical coupling in the anisotropic RSM ~\cite{Xie2020}.

In this paper, we study the quantum criticality around the critical coupling
of the isotropic RSM in the limit of $U\rightarrow \pm 1$ at finite
frequency ratio $\Delta /\omega $. The paper is structured as follows: In
Sec. II, we proposed the evidence of the second-order QPTs in the RSM at
finite frequency ratio by calculating the critical (dynamic) exponent of
energy gap analytically and analyze the singularity of ground state energy
numerically at $U=\pm 1$. We discuss the critical behavior of order
parameter and its finite size scaling behavior in Sec. III. In Sec. IV, we
propose the accurate hypothesis of fidelity susceptibility, and the nice
scaling behavior is also observed. We further explore the effects of the
truncation of the bosonic Hilbert space on the criticality in Sec. V, where
the comparison to the criticality of the QRM is also given. Finally we give
a summary and outlooks in Sec. VI. Appendix A provides the analytical
derivation of some critical exponents and Appendix B shows both finite-size
and finite-truncation analyses for the energy gap and the variance of
position quadrature of the field.

\section{Critical behavior in terms of the energy gap and the ground-state
energy}

As found by Xie ~\textsl{et al.} ~\cite{Xie2019}, the RSM at $U=1$ can be
mapped to an effective quantum harmonic oscillator. Thus the low energy
spectra of Hamiltonian (\ref{QRSM}) can be easily given by
\begin{equation}
\frac{\sqrt{-\left( \frac{\Delta }{2}+E\right) }\left( E+1-\frac{\Delta }{2}%
\right) }{\sqrt{-\left( \frac{\Delta }{2}+E+2g^{2}\right) }}=2n+1,\quad
n=0,1,2,...\infty .  \label{low}
\end{equation}%
where an infinite number of discrete energy levels is confined in the energy
interval
\begin{equation}
\frac{\Delta }{2}-1<E<E_{c}^{+},  \label{Interval}
\end{equation}%
if $g<g_{c}^{+}$, where $\ g_{c}^{+}=\sqrt{(1-\Delta )/2}$ and $%
E_{c}^{+}=-\Delta /2-2g^{2}$. All discrete low energy levels collapse to $%
E_{c}^{+}$ at $g=g_{c}^{+}$, thus the energy gap closes at $\ g_{c}^{+}$,
implying the occurrence of a second-order QPT in this model. For the case of
$U=-1$, the extension can be achieved straightforwardly by changing $\Delta $
into $-\Delta $. As the implementation of the RSM in trapped ions proposed
in Ref. ~\cite{Cong}, the Hamiltonian is derived in a rotated frame so that
the shifting detuning $\Delta $ \ can be negative.

For $g>g_{c}^{+}$, it is observed from numerical exact diagonalizations that
all energies become closer to $E_{c}^{+}$ monotonously with the increasing
truncated Fock space, although the convergence is hardly achieved by
numerics. We argue that the energy would be only $E_{c}^{+}=-\Delta
/2-2g^{2} $, and resulting effective harmonic potential, $\omega _{eff}=%
\sqrt{1+2g^{2}/\left( \frac{\Delta }{2}+E\right) }$, is flat. Note that the
energy could be any value for the flat potential. But the perquisite
condition for flat potential is no other than $E=E_{c}^{+}$. It is therefore
suggested that the gapless Goldstone mode excitations appear for $%
g>g_{c}^{+} $.

Below the critical points in the RSM, one can find that the ground-state has
a conserved parity symmetry, odd parity for $U=1$, and even parity for $U=-1$
~\cite{Xie2020}. Above the critical points, the parity symmetry is broken
due to the infinite degeneracy for all states.

The analytical mean photon number $\overline{n}=\left\langle a^{\dagger
}a\right\rangle $ in the ground-state is also found to diverge at $g_{c}^{+}$
~\cite{Xie2019}. Infinite photons are activated at the critical coupling,
signifying the emergence of a new quantum phase.

To corroborate the second-order QPT in the RSM, we then discuss critical
behavior of energy gap and singularity of the second derivative of the
ground-state energy.

As sown in Ref. \cite{Xie2020}, the energy gap between the ground-state and
the first excited state follow the universal scaling behavior in the
continuous QPT \cite{Sachdev}
\begin{equation}
\varepsilon (g\rightarrow g_{c})\sim \left\vert g-g_{c}\right\vert ^{z\nu
_{x}},
\end{equation}%
where $g_{c}$ is $g_{c}^{+}$ ($g_{c}^{-})$ for $U=1(-1)$, $\nu _{x}$ is the
critical exponent and $z$ is dynamical exponent. $z\nu _{x}$ is found to be $%
2$ as long as the counterrotating wave terms are present.

As given analytically in detail in Eq. (\ref{deltax}) of the Appendix A, the
variance of position quadrature of the field diverge as $\Delta x=\sqrt{%
\left\langle x^{2}\right\rangle -\left\langle x\right\rangle ^{2}}\propto
(g_{c}^{+}-g)^{-\frac{1}{2}}$. It suggests $\nu _{x}=\frac{1}{2}$, and thus
implies $z=4,$ in the RSM. In contrast, the gap exponent $z\nu _{x}=1/2$
with $\nu _{x}=1/4\ $and $z=2\ $has been found in both the Dicke model~\cite%
{Emary} and the QRM ~\cite{plenio2015}. Therefore the RSM is not in the same
universality class as the Dicke model and the QRM.

The phase transition takes places in the thermodynamic limit with the
emergence of singularity of some physical observables. Despite few degrees
of freedom in the QRM, the effective system size could be defined as $%
L_{R}=\Delta /\omega $ \cite{plenio2015} in order to catch sight of the
critical behavior of the phase transitions. The singularity only appears in
the limit of $\Delta /\omega \rightarrow \infty $. Since the energy gap
closes only at $U=\pm 1$ in the present RSM, we may define the system size
here as
\begin{equation}
L=\frac{1}{1\mp U}.  \label{L}
\end{equation}%
With this definition, $U=\pm 1$ is just corresponding to the thermodynamic
limit.

\begin{figure}[tbph]
\centering
\includegraphics[width=\linewidth]{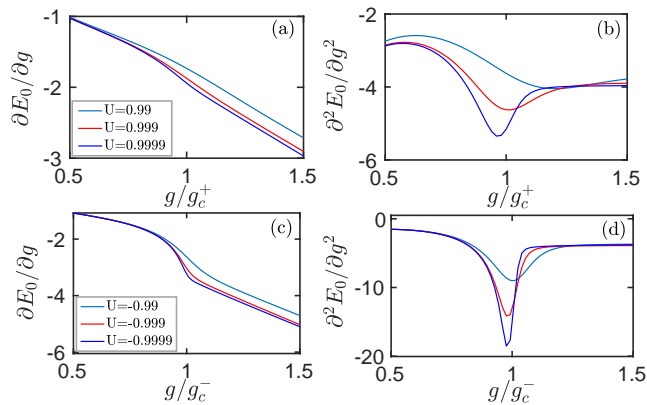}
\caption{ The first- and second-order derivative of the ground-state energy
by the numerically exact solution of the RSM at $\Delta =0.5$ for different $%
U$ approaching to $1$ (upper panel) and $-1$ (lower panel).}
\label{derivative}
\end{figure}

As long as $\left\vert U\right\vert <1$, the ground-state energy at
arbitrary coupling strength can be in principle calculated exactly. We then
look at its derivatives at the critical points $g_{c}^{\pm }=\sqrt{(1\mp
\Delta )/2}$ for finite effective sizes $L(U)$, and inspect how it evolves
with the increase of the system size. The exact solution can be obtained by
either the G-function approach ~\cite{Xie2019} or the direct exact
diagonalzations in truncated Fock space. Figure \ref{derivative} presents
the first-order (left panel) and the second-order \ (right panel)
derivatives of the ground state energy for $U$ gradually approaching to $\pm
1$, i.e. the size $L$ tends to infinity. One can find that the first-order
derivative is always continuous when $U\rightarrow \pm 1$, while its
second-order derivative changes drastically with system size $L$, displaying
the discontinuity in the trend. It thus provides the another evidence of the
second-order QPT in this model. It should be pointed out that it is
extremely difficult to obtain the converged results in the both approaches
if $U$ further approaches to $\pm 1$.

We will perform the finite-size scaling analysis \ on the order parameter
and fidelity by using the above definition of the effective system size
below.

\section{Order parameter and finite-size scaling analysis}

To characterize the continuous QPT, we should define the order parameter. In
QRM ~\cite{plenio2015}, the re-scaled cavity photon number $\overline{n}%
=\left\langle a^{\dagger }a\right\rangle $ can be regarded as the order
parameter since it is zero below and finite above the critical points. \ But
in the present RSM it is always finite below and diverges at the critical
point. \ We may define the inverse photon number $1/\overline{n}$ as the
order parameter $M$. It is finite below $g_{c}^{+}$, and tends to zero at $%
g_{c}^{+}$. Eq.(34) in Ref. \cite{Xie2019} in the limit of $g\rightarrow
g_{c}^{+}$ can be rewritten as
\begin{equation}
\overline{n}=\frac{6g^{2}\left( g_{c}^{+2}-g^{2}\right) }{E_{c}^{+}-E_{0}}+%
\frac{3}{16\left( g_{c}^{+2}-g^{2}\right) }\propto \left( g_{c}^{+}-g\right)
^{-1},  \label{ph}
\end{equation}%
Thus the exponent for order parameter is $\beta =1$. The detailed derivation
of Eq. (\ref{ph}) is also given in Eq. (\ref{meannumber}) in Appendix A.

\begin{figure}[tbp]
\centerline{\includegraphics[scale=0.4]{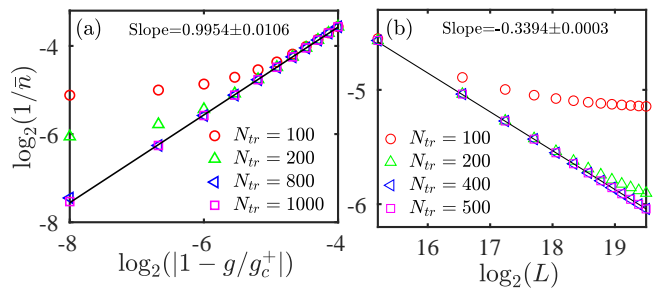}}
\caption{ (a) The log-log plot of the order parameter $1/\overline{n}$ as a
function of $\left|1-g/g_{c}^{+}\right|$ before $g_c$ for different
truncation photon numbers $N_{tr}$. The fitting slope is very close to 1. $%
U=1$ and $\Delta=0.5$. (b) The log-log plot of order parameter $1/\overline{n%
}$ as a function of system size $L$ at $g_c$. The fitting slope is very
close to $-1/3$. $\Delta=0.5$.}
\label{Np_log}
\end{figure}

To confirm the analytical findings, we perform the numerically
diagonalization of the RSM, which requires the truncation of the bosonic
Hilbert space. Figure ~\ref{Np_log} (a) shows the order parameter $1/%
\overline{n}$ as a function of the reduced coupling constant $t=1-g/g_{c}^{+}
$ on a log-log scale before $g_{c}^{+}$ by numerics for different truncation
photon numbers $N_{tr}$. One can find that the convergence can be achieved
for very large $N_{tr}$. A power law behavior, $1/\overline{n}\propto
|1-g/g_{c}^{+}|^{\beta },$ is gradually exhibited in a wider the critical
regime with increasing $N_{tr}$. The slope of the fitting line indicates
that  $\beta $  is just close to $1$. A log-log plot for the order parameter
versus the system size for different truncation photon numbers at the
critical point $g_{c}^{+}$ is also presented in Fig.~\ref{Np_log} (b). Note
that for finite size systems, i.e. $\ U$ is away from $1$, the convergence
is arrived at more quickly with increasing $\ N_{tr}$, compared to the
thermodynamic limit $\ U=1$. \ A perfect power-law behavior, $1/\overline{n}%
\propto L^{-\gamma }$, is clearly shown for the large truncation $N_{tr}$.
The fitted value of the scaling exponent $\gamma $ is very close to $1/3$.
Through extensive calculations, one can find that both $\beta $ and $\gamma $
are independent of the values of $\Delta $.

The finite-size scaling for the order parameter can be also performed. In
the critical regime of the continuous phase transition, it should satisfy
the finite-scaling ansatz for the homogeneous function $M(\lambda
^{h_{t}}t,\lambda ^{h_{L}}L)=\lambda M(t,L)$, where $\lambda $ is the scale
factor and $h_{t,L}$ related to some exponents below. Accordingly, the
homogeneous function takes the following form
\begin{equation}
{M}=|t|^{\beta }m(|t|L^{\gamma /\beta }),  \label{full_scale}
\end{equation}%
where $\beta $ is the order parameter exponent and\ $\gamma $ is the
finite-size scaling exponent for the order parameter. The function $m$
should be universal for large system size in the second-order QPT. $\nu
=\gamma /\beta $ is the correlation length exponent due to the scaling
function Eq. (\ref{full_scale}).

\begin{figure}[tbp]
\centerline{\includegraphics[scale=0.4]{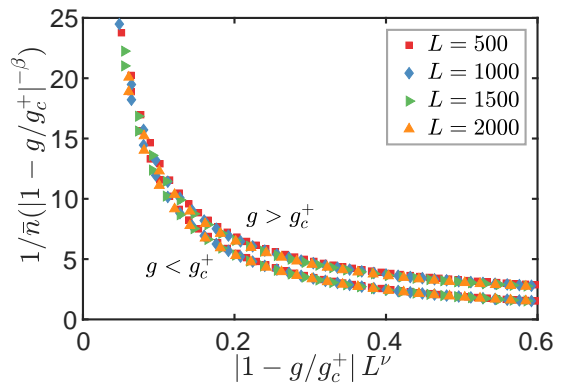}}
\caption{ Finite size scaling for order parameter $1/\overline{n}$ according
to Eq. (\protect\ref{full_scale}) for both above and below the critical
point at $U=1$ and $\Delta =0.5$.}
\label{Np_saling}
\end{figure}
By numerically diagonalizing the RSM, we show the finite-size scaling
according to Eq. (\ref{full_scale}), where $M$ is just the order parameter $%
1/\overline{n}$, for different values of $L$ above and below the critical
point in Fig.~\ref{Np_saling}. A very large truncation number ($N_{tr}=1000$%
) for photon space is used to ensure that all numerical results in the
finite size are converged. Here within the two exponents $\beta =1$ and $%
\gamma =1/3$ obtained above independently (cf. Fig.~\ref{Np_log}), an
excellent collapse to the single curve in the critical regime is achieved
for different large sizes. Two universal functions above and below the
critical points are not required to be the same ~\cite{plenio2015}. The
universal scaling behavior observed here corroborates the continuous QPTs in
the RSM at $U=1$. The correlation length exponent of the RSM is clearly $\nu
=1/3$ in terms of the length definition (\ref{L}).

\section{FIDELITY SUSCEPTIBILITY and finite-size scaling analysis}

The ground state fidelity can be taken as an useful tool to detect the QPTs
even without the knowledge of the order parameters~\cite{Gu2007,Gu2008}. It
is defined as the overlap of the ground state wavefunctions at very close
coupling parameters $g$ and $g+\delta g$ as
\begin{equation*}
F(g,\delta g)=\left\vert \left\langle \psi _{0}(g)|\psi _{0}(g+\delta
g)\right\rangle \right\vert .
\end{equation*}%
The singularity of critical point would inevitably leads to the wavefunction
experiencing dramatically change when crossing the critical point, showing a
sharp change of fidelity.

Correspondingly, the fidelity susceptibility can be defined as~\cite{Gu2007}
\begin{equation*}
\chi _{F}(g)=\lim_{\delta g\rightarrow 0}\frac{-2\ln F(g,\delta g)}{\delta
g^{2}},
\end{equation*}%
and can be written in terms of the eigenstates of the Hamiltonian
\begin{equation}
\chi _{F}(g)=\sum_{n\neq 0}\frac{\left\vert \left\langle \psi
_{n}(g)\right\vert H_{I}\left\vert \psi _{0}(g)\right\rangle \right\vert ^{2}%
}{\left[ E_{n}(g)-E_{0}(g)\right] ^{2}},
\end{equation}%
which is the leading order of the change in the fidelity.

In the critical regime, the fidelity susceptibility decays in a power law
away from the critical point with critical exponent $\alpha $ ~\cite%
{Gu2008,Kwok2008}
\begin{equation}
\chi _{F}\propto |g-g_{c}^{\pm}|^{-\alpha }.  \label{chiFgc}
\end{equation}

In terms of the original definition of fidelity susceptibility \cite{Gu2010}
\begin{equation}
\chi_F=\left<\frac{\partial\Psi_{0}}{\partial \lambda} \bigg|\frac{\partial
\Psi_{0}}{\partial \lambda} \right>-\left<\frac{\partial \Psi_{0}}{\partial
\lambda}\bigg|\Psi_{0} \right>\left<\Psi_{0} \bigg|\frac{\partial \Psi_{0}}{%
\partial \lambda}\right>,  \label{chi1}
\end{equation}
we can actually derive the critical exponent of fidelity susceptibility $%
\alpha =2$ analytically, which is described in the end of Appendix A.

\begin{figure}[tbph]
\centering
\includegraphics[width=\linewidth]{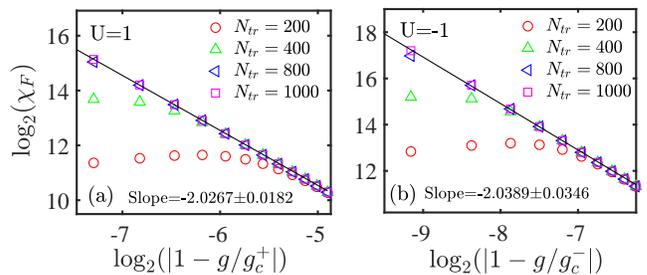}
\caption{ The log-log plot of fidelity susceptibility as a function of $%
g_{c}^{\pm }-g$ for different truncation photon numbers $N_{tr}$ at $U=1$
(left) and $-1$ (right). $\Delta =0.5$.}
\label{fidelity}
\end{figure}

Similar to the order parameter, we present the fidelity susceptibility as a
function of the deviation $g_{c}-g$ on a log-log scale below $g_{c}$ for $%
U=\pm 1$ by numerics within truncation of the bosonic Hilbert space in Fig. %
\ref{fidelity}. Converged results are convincingly obtained with large
truncation numbers. The converged data can be well fitted by straight lines
with the same slope very close to $-2$, so we can immediately obtain its
critical exponent $\alpha =2$, consistent with the analytical derivation.

In the second QPTs, the average fidelity susceptibility should follow the
following finite-scaling ansatz \cite{Gu2007,Liu2009}
\begin{equation}
\frac{\chi _{F}^{max}-\chi _{F}(g)}{\chi _{F}(g)}=f[(g-g_{m})L^{\nu }],
\label{FS_uni}
\end{equation}%
where $g_{m}$ is the coupling strength of the maximum fidelity
susceptibility, $f$ is the scaling function, and $\nu $ is the correlation
length critical exponent.

\begin{figure}[tbph]
\centering
\includegraphics[width=\linewidth]{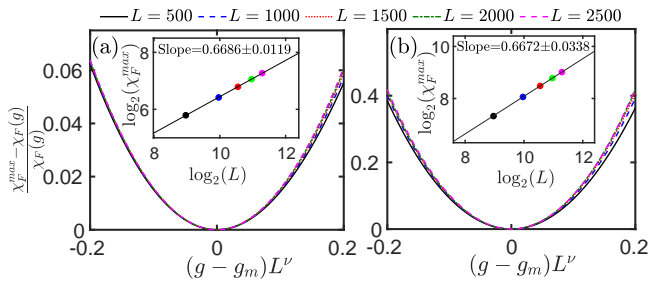}
\caption{ The finite size scaling for the fidelity susceptibility according
to Eq. (\protect\ref{FS_uni}) for different system size $L$ for $%
U\rightarrow 1$ (a) and $U\rightarrow -1$ (b). $\Delta =0.5$. The
corresponding size $L$ dependence of the maximum fidelity susceptibility $%
\protect\chi _{F}^{max}$ is presented in the insets.}
\label{susceptibility}
\end{figure}

This function should be universal for large $L$ in the second-order QPTs. As
shown in Fig. \ref{susceptibility}, with $\nu =1/3$, an excellent collapse
in the critical regime is achieved according to Eq. (\ref{FS_uni}) in the
curves for large effective sizes $L$. Beyond the critical regime, the
collapse becomes poor. As $L$ increases, the curves tend to converge in the
wider coupling regime. It is demonstrated that $\nu $ is an universal
constant. Interestingly, the correlation length exponent obtained here is
consistent with that obtained from the finite-size scaling ansatz in the
order parameter in the last section. The average FS diverges as the system
size increases at the critical point, convincingly demonstrating a
Landau-type phase transition in the RSM.

The insets of Fig. \ref{susceptibility} show fidelity susceptibility at the
maximum point as a function of the system size $L$ for both $U\rightarrow
\pm 1$ on a log-log scale. A nice power law behavior $\chi _{F}^{max}\propto
L^{\mu }$ is shown in both cases. The scaling exponent $\mu $ fitted from
the numerical data at large $L$ tends to $2/3$.

In terms of the finite size scaling ansatz in Eq. (\ref{FS_uni}), the
critical exponents should satisfy the scaling law as $\mu /\nu =\alpha $,
where $\alpha$ is just the fidelity susceptibility critical exponent.
Obviously, $\alpha =2$ here is the same as the independent calculation given
in Fig. \ref{fidelity}, as well as the analytical derivation in the Appendix
A.

\section{Finite-Truncation Effects on the criticality}

The numerical solution of the RSM model requires one additional finite size
parameter, namely the truncation of the bosonic Hilbert space. Does this
truncation lead to additional critical exponents? In this section, we will
demonstrate the solely effect of the photonic truncation $N_{tr}$ on the
criticality of the RSM at $U=\pm 1$ by directly diagonalizing the RSM.

\begin{figure}[tbph]
\centering
\includegraphics[width=\linewidth]{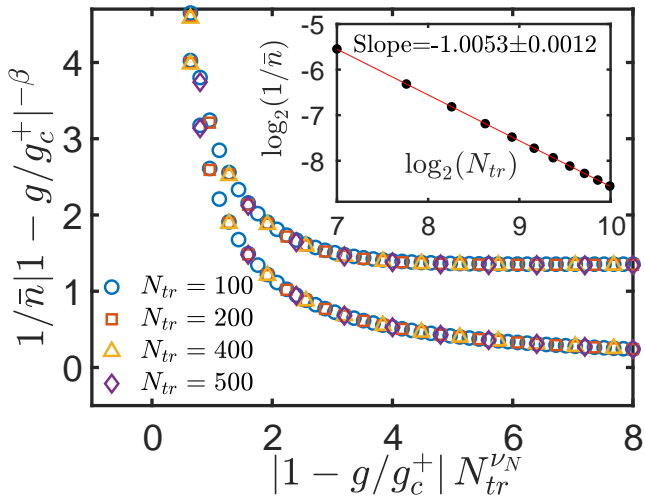}
\caption{(a) Finite-size effects for the order parameter $1/\bar{n}$ in
terms of the Hilbert space truncation according to Eq.~(\protect\ref%
{scalingNtr}) for both above and below the critical point. Inset: the
corresponding log-log plot of $N_{tr}$ and $1/\bar{n}$ with fitting slope
close to $-1$. }
\label{OPscalingNtr}
\end{figure}

Using $N_{tr}$ as the alternative finite-size parameter, we first perform
the finite-truncation scaling analysis on the RSM for the order parameter $1/%
\bar{n}$, and calculate the critical scaling exponents. Inset of Fig.~\ref%
{OPscalingNtr} presents the order parameter $1/\overline{n}$ as a function
of the photon truncation $N_{tr}$ right at the critical coupling strength on
a log-log scale by numerics. \ A nice power law behavior satisfying $1/\bar{n%
}\propto N_{tr}^{-\gamma _{N}}$ with $\gamma _{N}$ very close to $1$ is
demonstrated. Interestingly, the critical scaling exponent in terms of the
bosonic truncation number is different from that $\gamma =1/3$ in terms of
the effective system size $L$ defined in (\ref{L}) found in the Sec. III. By
comparing the two size exponents of the order parameter, it is quite
intriguing to note that two effective sizes can be related through $%
N_{tr}\propto L^{1/3}$.

The finite-truncation scaling function for order parameter is just similar
to Eq.~(\ref{full_scale}) by taking $N_{tr}$ as the system size as following
\begin{equation}
M=|1-g/g_{c}^{+}|^{\beta }f_{N}(|1-g/g_{c}^{+}|N_{tr}^{{\gamma _{N}/\beta }%
}),  \label{scalingNtr}
\end{equation}%
Correspondingly, $\nu _{N}=\gamma _{N}/\beta $ denotes the correlation
length exponent in terms of the bosonic truncation number $N_{tr}$ due to
the scaling function. By numerics, we can also present the finite-truncation
scaling according to Eq.~(\ref{scalingNtr}) for different values of $N_{tr}$
above and below the critical point at $U=1$ in Fig.~\ref{OPscalingNtr}.
Employing two exponents $\beta =1$ and $\gamma _{N}=1$ obtained
independently above, we also observe that the scaling functions for
different $N_{tr}$ collapse to the single curve, exhibiting an perfect
universal scaling behavior. It follows that a universal critical behavior
can be also established in terms of the bosonic Hilbert space truncation
number, but leading to an alternative critical correlation length exponent
of the RSM, $\nu _{N}=1$.

\begin{figure}[tbph]
\centering
\includegraphics[width=\linewidth]{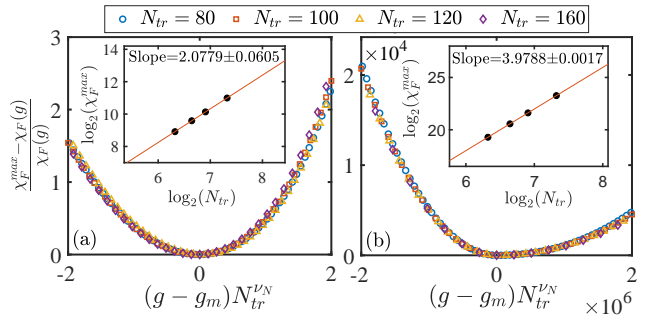}
\caption{Finite-size effects for the fidelity susceptibility of the RSM and
the quantum Rabi model in terms of Hilbert space truncation. Inset: the
corresponding log-log plot of $N_{tr}$ and $\protect\chi _{F}^{max}$ with
fitting slope close to $2$. $U=1,\Delta =0.5$.}
\label{SFPscalingNtr}
\end{figure}

Then we turn to the finite-truncation scaling of the average fidelity
susceptibility using $N_{tr}$. The corresponding scaling ansatz can be given
by Eq. (\ref{FS_uni}) if $L$ is replaced by $N_{tr}$. Figure ~\ref%
{SFPscalingNtr} (a) show excellent universal scaling for large $N_{tr}$ with
the correlation length exponent $\nu _{N}=1$. The inset exhibits a very good power
law behavior $\chi _{F}^{max}\propto N_{tr}^{\mu _{N}}$ with $\mu _{N}=2$.
These scaling exponents are also different from those in terms the size $L$
defined in Eq. (\ref{L}).

\begin{center}
\begin{table}[h]
\caption{ Summary of the critical exponents and the critical scaling
exponents in the Rabi-Stark model. The critical exponent:  $A\propto
\left\vert g-g_{c}\right\vert ^{y}$, the  scaling exponents: $%
A\propto L^{y_{L}}$ in $L$, and  $A\propto
\left( N_{tr}\right) ^{y_{N}}$ in $N_{tr}$, respectively.}%
\begin{tabular}{p{2cm}p{1.5cm}p{1.50cm}p{1.5cm}p{1.5cm}p{-0.50cm}}
\hline\hline
exponent & $\varepsilon$ & $1/\overline{n}$ & $\Delta x$ & $\chi _{F}$ &  \\
\hline
$y$ & 2 & 1 & -1/2 & -2 &  \\
$y_L$ & -2/3 & -1/3 & 1/6 & 2/3 &  \\
$y_N$ & -2 & -1 & 1/2 & 2 &  \\ \hline\hline
\end{tabular}%
\end{table}
\end{center}

At this stage, we may collect critical exponents and critical scaling
exponents for more observables for completeness, which are summarized in
Table I. The scaling exponents for the energy gap $\varepsilon $ and the
variance of position quadrature of the field $\Delta x$ are calculated in
the Appendix B.  For an unified description,  the critical exponent is defined by $A\propto
\left\vert g-g_{c}\right\vert ^{y}$, the finite-size scaling exponents by $%
A\propto L^{y_{L}}$, and the finite-truncation scaling exponent by $A\propto
\left( N_{tr}\right) ^{y_{N}}$, respectively, where $y$'s stand for the corresponding
exponents of the observables $A$. One can note that all ratios of two scaling exponents $y_{L}/y_{N}$ are $1/3$,
providing the consistent evidence for $N_{tr}\propto L^{1/3}$. The
correlation length exponents $\nu =1/3$ in $L$ and $\nu _{N}=1$ in $N_{tr}$
can be seen by the ratios between the critical exponent and the scaling
exponent, i.e.  $y/y_{L}$ and $y/y_{N}$, for all observables.

Finally, for comprehensive comparisons, we revisit the criticality of the
QRM by exploring the effects of the Hilbert space truncation. For brevity,
we only analyze the finite-truncation scaling of the average fidelity
susceptibility, which is independent of the order parameter. To facilitate
the analysis of the criticality of the QRM, we set $L_{R}=\Delta /\omega
=10^{8}$. In this case, the critical coupling of the QPT \cite{plenio2015}
is around $\allowbreak 5000\omega $. Fig.~\ref{SFPscalingNtr} (b) show
excellent universal scaling for large $N_{tr}$ with the correlation length
exponent $\nu _{N}^{R}=2$ for the QRM. The inset show a very good power law
behavior $\chi _{F}^{max}\propto N_{tr}^{\mu _{N}^{R}}$ with $\mu _{N}^{R}=4$%
. Very importantly, we can find that in terms of the same Hilbert space
truncation, the critical correlation length exponent for the RSM is
different from that in the QRM, indicating the different universality class.

Interestingly, comparing to the previous critical correlation length
exponent $\nu^R =2/3$ in the QRM by the universal scaling of the same
fidelity susceptibility using $L_{R}=\Delta /\omega $ ~\cite{Wei2018}, it is
quite intriguing to note that two sizes in the QRM can be also related
through $N_{tr}\propto L_{R}^{1/3}$. Comparing the effects of the Hilbert
space truncations on both the RSM and QRM, we can see that the length
defined in Eq. (\ref{L}) of the RSM plays the equivalent role as the
effective length $L_{R}=\Delta /\omega $ in the QRM. However, with these two
equivalent length scales, the critical correlation length exponent $\nu =1/3$
of the RSM obtained in the last section is also different from $\nu^R =2/3$
in the QRM.

Using the atomic number, the finite-size scaling analysis for the fidelity
susceptibility has been performed for the Dicke model ~\cite{Liu2009}, and
the correlation length exponent is obtained to be $2/3$, which is the same
as that in Lipkin-Meshkov-Glick model \cite{Kwok2008} and the QRM using the
effective size $L_{R}=\Delta /\omega $ \cite{Wei2018}. It is now generally
accepted that the effective size $L_{R}=\Delta /\omega $ in the QRM plays
the same role on the criticality as the atomic numbers of the Dicke model
and the Lipkin-Meshkov-Glick model. No any different critical behavior
of the continuous QPT has been observed in the Dicke model,
Lipkin-Meshkov-Glick model, as well as the QRM in the infinite frequencies
ratio. However, we indeed find that the critical correlation length exponent
in the RSM is truly exceptional to all these models. Therefore, we believe
that the RSM is in a different universality class.

\section{Summary}

We reveal the second-order QPT in the RSM at $U=\pm 1$ by studying the
critical behavior of the energy gap, the order parameter, and the fidelity
in this paper. The critical exponents for several observables can be
analytically obtained. The energy gap exponent is $z\nu _{x}=2$ with $\nu
_{x}=1/2$, in sharp contrast to $z\nu _{x}=1/2$ with $\nu _{x}=1/4$ in both
the Dicke model~ and the QRM. The critical exponent of the inverse photon
number, which can be regarded as the order parameter of the RSM, is $\beta =1
$, and the critical exponent of fidelity susceptibility is $\alpha =2 $. All
these critical exponents have also been confirmed by the numerically
diagonalizing the RSM.

By using the effective system size $L=1/\left( 1\mp U\right) $ for $%
U\rightarrow \pm 1$, the finite size scaling analyses for both the order
parameter and the fidelity susceptibility are performed, and the universal
scaling behaviors are clearly demonstrated. The size scaling exponent of the
order parameter is found to be $\gamma =1/3$. Two size exponents of the
fidelity susceptibility are also obtained by the perfect finite size
scaling, which ratio recover the size independent critical exponent of the
fidelity susceptibility $\alpha =2$. The finite size scaling analyses for
the energy gap $\varepsilon $ and the variance of position quadrature of the
field $\Delta x$ are also presented. The critical correlation exponent $\nu
=1/3$ is obtained consistently in the calculation of all observables. A
consistent picture is achieved and provides a strong evidence of the
second-order QPT in the RSM.

We also explore the effects of the truncation of the bosonic Hilbert space
on the criticality of the RSM. The finite-truncation scaling analyses for
the several observables including the order parameter and the fidelity
susceptibility are performed. \ Alternative critical scaling exponents in
terms of the photonic truncation number $N_{tr}$ are found. The universal
critical behavior can be also established in terms of the truncation number.
In particular, the corresponding critical correlation length exponent is $%
\nu _{N}=1/3$ in the RSM. Comparing with the scaling exponents obtained in
the effective system size $L $, we can find $N_{tr}\propto L^{1/3}$.

To compare to the criticality of the QRM at infinite frequency ratio on the
same foot, we also perform the finite-truncation scaling analysis for the
order parameter independent fidelity susceptibility in the QRM. We find that
the critical correlation length exponent in the RSM is different from $\nu
_{N}^{R}=2/3$ in the QRM, suggesting that these two models would belong to
different universality classes.

We like to point out that the QPT found in the RSM is of practical
importance in two aspects. \ One is that we do not need the extreme
condition for the occurrence of the second-order QPT like the infinite ratio
of the qubit and cavity frequencies $\Delta /\omega $ in the QRM. \ In the
present RSM, the frequency ratio $\Delta /\omega $ allowed for phase
transitions can be any values. The other one is that for $U $ identical to
the cavity frequency, the critical coupling is $g_{c}^{+}=\sqrt{(1-\Delta )/2%
}$, which can be extremely weak near the resonance by tuning the qubit
frequency. This QPT should be experimentally feasible in any cavity-atom
coupling device, such as the cavity (circuit) QEDs and the ion-trap. The
price that should be paid is to incorporate the nonlinear Stark coupling
between the two-level system and the fields in the devices.

\textbf{ACKNOWLEDGEMENTS} This work is supported by the National Science
Foundation of China (Nos. 11674285, 11834005), the National Key Research and
Development Program of China (No. 2017YFA0303002).

$^{\ast }$ Corresponding author. Email:qhchen@zju.edu.cn.

\begin{appendix}

\section{Analytical derivation of some critical exponents of the RSM at $U=1$%
}

In this Appendix, we derive some critical exponent analytically based on the
analytical exact solutions to the RSM at $U=1$. We will first briefly review
the previous solution in the framework of the Fock space \cite{xieCTP}, then
we present the exact eigenfunction.

In terms of the basis of $\sigma _{z}$, the Hamiltonian (\ref{QRSM}) in the
matrix form can be written as
\begin{equation}
H_{0}=\left(
\begin{array}{cc}
2a^{\dag }a+\frac{\Delta }{2} & g\left( a^{\dag }+a\right)  \\
g\left( a^{\dag }+a\right)  & -\frac{\Delta }{2}%
\end{array}%
\right) .
\end{equation}%
The series expansion of the eigenfunction in the Fock space is
\begin{equation}
\left\vert \Psi \right\rangle =\binom{\left\vert \Psi _{1}\right\rangle }{%
\left\vert \Psi _{2}\right\rangle }=\left(
\begin{array}{c}
\sum_{n}^{\infty }e_{n}\left\vert \phi _{n}\right\rangle  \\
\sum_{n}^{\infty }f_{n}\left\vert \phi _{n}\right\rangle
\end{array}%
\right) .
\end{equation}%
where $e_{n}$ and $f_{n}$ are the expansion coefficients, $\left\vert \phi
_{n}\right\rangle $ is the number states. The Schr\"{o}dinger equation then
gives
\begin{eqnarray}
&&\left( 2a^{\dag }a+\frac{\Delta }{2}\right) \sum_{n}^{\infty
}e_{n}\left\vert \phi _{n}\right\rangle +g\left( a^{\dag }+a\right)
\sum_{n}^{\infty }f_{n}\left\vert \phi _{n}\right\rangle   \notag \\
&=&E\sum_{n}^{\infty }e_{n}\left\vert \phi _{n}\right\rangle ,  \label{s1}
\end{eqnarray}%
\begin{equation}
g\left( a^{\dag }+a\right) \sum_{n}^{\infty }e_{n}\left\vert \phi
_{n}\right\rangle =\left( \frac{\Delta }{2}+E\right) \sum_{n}^{\infty
}f_{n}\left\vert \phi _{n}\right\rangle .  \label{s2}
\end{equation}%
By inspection of Eq.~(\ref{s2}), we obtain
\begin{equation*}
\sum_{n}^{\infty }f_{n}\left\vert \phi _{n}\right\rangle =\frac{g\left(
a^{\dag }+a\right) }{\frac{\Delta }{2}+E}\sum_{n}^{\infty }e_{n}\left\vert
\phi _{n}\right\rangle ,
\end{equation*}%
Inserting it into Eq. (\ref{s1}) leads to the effective Hamiltonian for the
bosonic wavefunction in the upper level $\left\vert \Psi _{1}\right\rangle $%
%
\begin{equation}
H_{eff}=2a^{\dagger }a+\chi (a^{\dagger }+a)^{2}+\frac{\Delta }{2},
\end{equation}%
with $\chi =\frac{g^{2}}{\frac{\Delta }{2}+E}$.

To diagonalize the Hamiltonian above, we apply the squeezing transformation $%
S=\exp {[\frac{r}{2}(a^{2}-{a^{\dagger }}^{2})]}$ with $r=\frac{1}{4}\ln (%
\frac{1}{1+2\chi })$, and then get a Hamiltonian for a quantum oscillator
\begin{equation}
H^{\prime }=SH_{eff}S^{\dagger }=\sqrt{1+2\chi }(2a^{\dagger }a+1)-1+\frac{%
\Delta }{2}.
\end{equation}%
The eigenenergy $E$ is obviously given by
\begin{equation}
E=\sqrt{1+2\chi }(2n+1)-1+\frac{\Delta }{2},\quad n=0,1,2...\infty.
\end{equation}%
which further results in Eq.~(\ref{low}).

The eigenfunction to the harmonic oscillator is just the Fock state $%
\left\vert n \right\rangle$, thus the eigenfunction for the RSM at $U=1$ for
the low spectra reads
\begin{equation}
\left\vert \Psi _{n}\right\rangle =\frac{1}{N_{n}}%
\begin{pmatrix}
c_{n}S^{\dagger }\left\vert n\right\rangle \\
d_{n}S^{\dagger }(a^{\dagger }+a)\left\vert n\right\rangle%
\end{pmatrix}
\label{wavefun}
\end{equation}%
where $c_{n}=(1+2\chi _{n})^{1/4}$ and $d_{n}=\chi _{n}/g$, and the
normalization factor $N_{n}=\sqrt{c_{n}^{2}+(2n+1)d_{n}^{2}}$. Several
quantities can be calculated using this eigenfunction. For convenience, we
denote the dimensionless coupling parameter $\lambda =g/g_{c}^{+}$.

First, we calculate the mean photon number using the wave function of Eq.~(%
\ref{wavefun}) as
\begin{eqnarray}
\left\langle a^{\dagger }a\right\rangle &=&\frac{1}{N_{n}^{2}}%
[c_{n}^{2}\left\langle n\right\vert Sa^{\dagger }aS^{\dagger }]\left\vert
n\right\rangle  \notag \\
&&+d_{n}^{2}\left\langle n\right\vert (a^{\dagger }+a)Sa^{\dagger
}aS^{\dagger }(a^{\dagger }+a)\left\vert n\right\rangle ]  \notag \\
&=&\frac{1}{N_{n}^{2}}\{c_{n}^{2}(n\cosh 2r+\sinh ^{2}r)  \notag \\
&&+d_{n}^{2}[\left( 2n^{2}+n+1\right) \cosh 2r+(2n+1)\sinh ^{2}r  \notag \\
&&+n(n+1)\sinh 2r]\}  \notag \\
&=& \frac{3}{8}\frac{(2n+1)^{2}+1}{(1-\Delta )(1-\lambda ^{2})}-\frac{1}{2} %
\left[1-(1-\Delta ) \lambda ^2\right]  \notag \\
&\simeq &\frac{3}{8}\frac{(2n+1)^{2}+1}{(1-\Delta )(1-\lambda ^{2})}\propto
(1-\lambda )^{-1}.  \label{meannumber}
\end{eqnarray}%
where the limit $g\rightarrow g_{c}$, i. e. $\lambda$ is a small quantity,
is performed in the last step. This is just Eq.~(\ref{ph}) for the order
parameter in the main text.

Then we turn to the variance of position quadrature of the field $\Delta x$.
By using $x=1/\sqrt{2}(a^{\dagger }+a)$ , we have
\begin{eqnarray}
(\Delta x)^2 &=&\left\langle x^{2}\right\rangle -\left\langle x\right\rangle
^{2}=\frac{1}{2}\left\langle (a^{\dagger }+a)^{2}\right\rangle  \notag \\
&=&\frac{e^{2r}}{2N_{n}^{2}}\left[ c_{n}^{2}(2n+1)+d_{n}^{2}(6n^{2}+6n+3)%
\right]  \notag \\
&\simeq &\frac{3\left[ (2n+1)^{2}+1\right] }{4(1-\Delta )\left( 1-\lambda
^{2}\right) }\propto (1-\lambda )^{-1}.  \label{deltax}
\end{eqnarray}%
The critical exponent $\nu_x=1/2$ for the position fluctuation is obtained,
which is part of the gap exponent in the main text. The relative difference
for the mean photon number and the position is within $2\%$ at $\lambda=0.9$.

Finally we calculate the ground state fidelity susceptibility $\chi _{F}$ in
the limit of $g\rightarrow g_{c}^{+}$, by using its the differential form
Eq. (\ref{chi1}). The eigenfunction of n-th energy level in this limit can
be written as
\begin{equation}
\left\vert \Psi _{n}\right\rangle =%
\begin{pmatrix}
C_{n}S^{\dagger } \left\vert n \right\rangle \\
D_{n}S^{\dagger }(a^{\dagger }+a)\left\vert n \right\rangle%
\end{pmatrix}
,
\end{equation}%
where
\begin{eqnarray*}
C_n=\frac{c_n}{N_n}&=&\frac{c_n}{\sqrt{c_n^2+(2n+1)d_n^2}}  \notag \\
&=& \frac{\sqrt{1+2\chi}}{\sqrt{\sqrt{1+2\chi}+(2n+1)\chi^2/g^2}}  \notag \\
&\simeq & \frac{(1-\Delta)\lambda}{2n+1}\sqrt{2(1-\lambda^2)} \\
D_n=\frac{d_n}{N_n}&=&\frac{d_n}{\sqrt{c_n^2+(2n+1)d_n^2}}  \notag \\
&=& \frac{\chi/g}{\sqrt{\sqrt{1+2\chi}+(2n+1)\chi^2/g^2}}  \notag \\
&\simeq & \sqrt{\frac{1}{2n+1}}
\end{eqnarray*}

Now, we can calculate the first-order partial derivative of the ground state
wave function with respect to $\lambda $
\begin{equation*}
\left\vert \frac{\partial \Psi _{0}}{\partial \lambda }\right\rangle =%
\begin{pmatrix}
(\frac{\partial C_{0}}{\partial \lambda })S^{\dagger }+C_{0}(\frac{\partial
S^{\dagger }}{\partial \lambda })\left\vert 0 \right\rangle \\
D_{0}(\frac{\partial S^{\dagger }}{\partial \lambda })(a^{\dagger
}+a)\left\vert 0 \right\rangle%
\end{pmatrix}%
,
\end{equation*}%
where
\begin{equation}
\frac{\partial S^{\dagger }}{\partial \lambda }=-\frac{1}{2}\frac{\partial
r(\lambda )}{\partial \lambda }S^{\dagger }(a^{2}-{a^{\dagger }}^{2})
\end{equation}%
\begin{equation}
\frac{\partial r}{\partial \lambda }=-\frac{1}{2(1+2\chi )}\frac{\partial
\chi }{\partial \lambda }\simeq \frac{\lambda }{1-\lambda ^{2}}.
\end{equation}%
Some overlaps in Eq. (\ref{chi1}) are list below after lengthy calculation
\begin{eqnarray}
\left\langle \frac{\partial \Psi _{0}}{\partial \lambda }\bigg|\frac{%
\partial \Psi _{0}}{\partial \lambda }\right\rangle &=& \left( \frac{%
\partial C_0}{\partial\lambda}\right)^2+\frac{C_0^2}{2}\left( \frac{\partial
C_0}{\partial\lambda}\right) +\frac{3 D_0^2}{2}\left( \frac{\partial C_0}{%
\partial\lambda}\right)^2  \notag \\
&\simeq & \frac{(1-\Delta)^{2}\left( 2+\lambda ^{4}\right) }{\left(
1-\lambda ^{2}\right) }+\frac{3\lambda ^{2}}{2\left( 1-\lambda ^{2}\right)
^{2}}, \\
\left\langle \Psi _{0}\bigg|\frac{\partial \Psi _{0}}{\partial \lambda }%
\right\rangle &=& \left\langle \Psi _{0}\bigg|\frac{\partial \Psi _{0}}{%
\partial \lambda }\right\rangle = C_0 \frac{\partial C_0}{\partial \lambda}
\notag \\
&\simeq& -{2(1-\Delta )^{2}\lambda }.
\end{eqnarray}

We now arrive at the fidelity susceptibility in the limit to the critical
point
\begin{eqnarray}
\chi _{F} &\simeq &\frac{(1-\Delta )^{2}\left( 2+\lambda ^{4}\right) }{%
\left( 1-\lambda ^{2}\right) }+\frac{3}{2}\frac{\lambda ^{2}}{%
\left(1-\lambda ^{2}\right) ^{2}} +4(1-\Delta)^2\lambda^2  \notag \\
&\simeq &\frac{3}{2}\frac{\lambda ^{2}}{\left(1-\lambda ^{2}\right) ^{2}}
\propto (1-\lambda )^{-2},  \label{chif2}
\end{eqnarray}%
which explicitly yields the critical exponent for the fidelity
susceptibility as $\alpha =2$. The relative difference is within $4\%$ at $%
\lambda=0.99$.

\section{ Finite-size  scaling   for the energy gap $\varepsilon$  and the variance of position quadrature of the field  $\Delta x$}

The finite-size scaling function near critical point $g_{c}^{+}$ at $U=1$
for any observable $A$ should satisfy following form
\begin{equation}
A(\mathbb{L},g)=|1-g/g_{c}^{+}|^{\beta }f(|1-g/g_{c}^{+}|\mathbb{L}^{\nu
^{\prime }})  \label{A_ScalFunc}
\end{equation}%
where $\nu ^{\prime }$ is the critical correlation length exponent, the size
$\mathbb{L}$ could be either the effective system size $L$ defined in Eq. (%
\ref{L}) or the Herbert space truncation $N_{tr}$. Accordingly, the
critical exponent $\nu ^{\prime }$ is either $\nu $ in terms of $L$  or $\nu
_{N}$ in $N_{tr}$.

\begin{figure}[tbph]
\centering
\includegraphics[width=\linewidth]{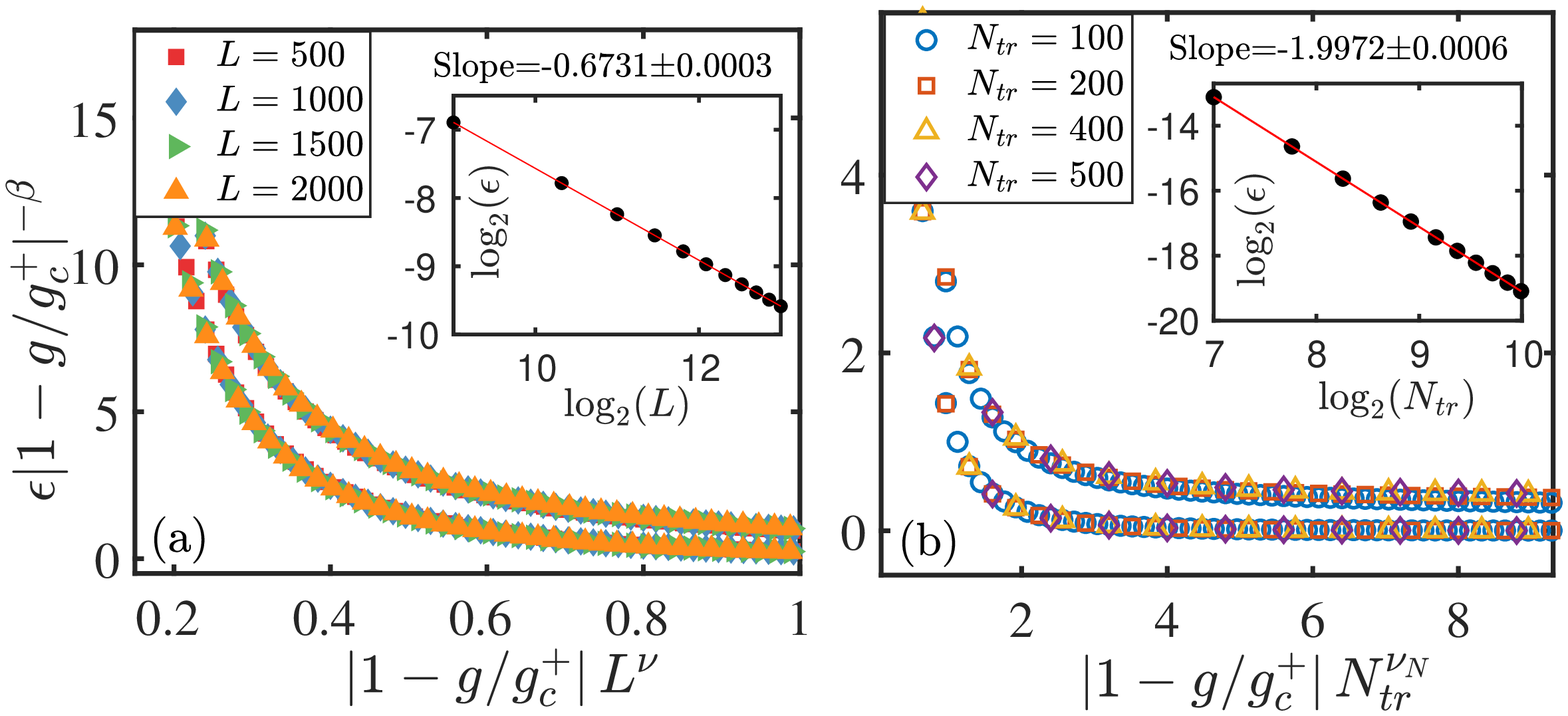}
\includegraphics[width=\linewidth]{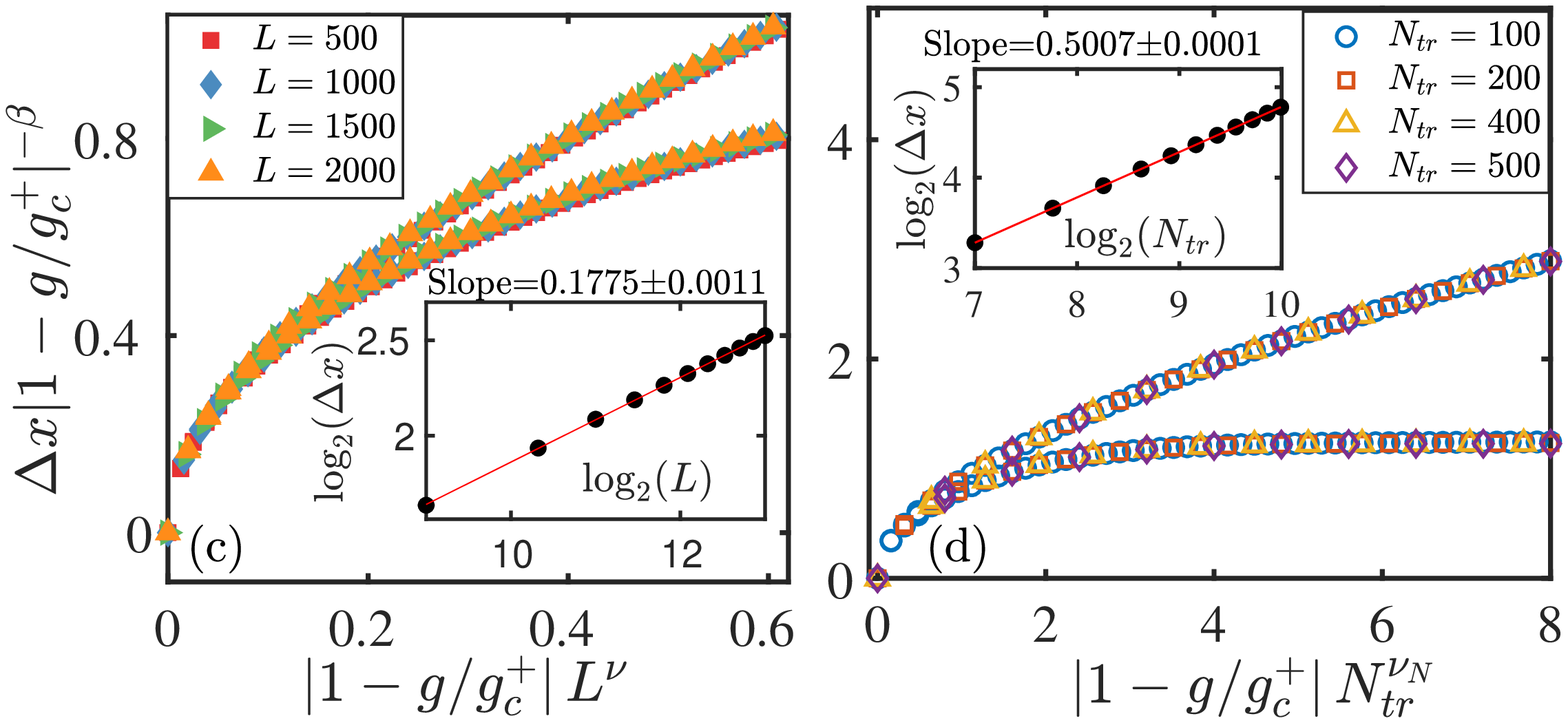}
\caption{Finite size scaling for the energy gap $\protect\epsilon$ (upper panel) and the variance of position quadrature of
the field $\Delta x$ (lower panel) according
to Eq.~(\protect\ref{A_ScalFunc}) in terms of $\mathbb{L}$  for both above and below the critical
point. The corresponding inset displays  the  log-log plot of $\protect\epsilon$ ($\Delta x$ ) and $\mathbb{L}$.  $\mathbb{L}=L$ for  left panel and   $\mathbb{L}=N_{tr}$ at $U=1$ for right panel. $\Delta=0.5.$}
\label{ep_scaling}
\end{figure}

Here we analyze the finite-size scaling for the energy gap $\epsilon $ and
the variance of position quadrature of the field $\Delta x\ $\ in terms of both
the effective system size $L$   and the bosonic Hilbert space  truncation $N_{tr}$  in
Fig.~\ref{ep_scaling}  by diagoanlizing the RSM. \
The insets of the upper panel of Fig.~\ref{ep_scaling} show the energy gap $\epsilon $ as a
function of $\mathbb{L}$ on a log-log scale by numerics. \ A nice power law
behavior is  observed with slopes very close to $-2/3$ and $-2$, respectively. The insets
of the lower  panel of Fig.~\ref{ep_scaling} is plotted for the variance of position quadrature of the
field $\Delta x$, and the power law behavior gives the slops $1/6$ and $1/2$
in two sizes, respectively. For  $\epsilon $, with the critical exponent
obtained in the main text $\beta =2$ and the critical correlation length
exponent $\nu ^{\prime }=1/3$ and $1$ for the two size, the excellent
collapse can be find in the upper panel of Fig.~\ref{ep_scaling}. Similarly, for $\Delta x$,
the critical exponent obtained in the main text $\beta =-1/2$, using the
same values of $\nu ^{\prime }$ in the upper panel of Fig.~\ref{ep_scaling}, we also note the perfect
collapse in the lower panel of Fig.~\ref{ep_scaling}.

\end{appendix}

\end{document}